\documentclass[runningheads]{llncs}
\usepackage{graphicx}
\usepackage{wrapfig}
\usepackage{tabularray}
\usepackage{color}
\usepackage{subcaption}
\usepackage{amsmath}
\usepackage[switch]{lineno}
\begin{document}
%

\title{Blockchain-based vs. SQL Database Systems for Digital Twin Evidence Management: A Comparative Forensic Analysis}
\titlerunning{Blockchain-based vs. SQL Database Systems}
%
\author{Boyd Franken \and
Hong-Hanh Nguyen-Le \and
Nhien-An Le-Khac\thanks{Corresponding author}}
\authorrunning{B. Franken et al.}
%
\institute{University College Dublin, Dublin, Ireland \\
\email{\{boyd.franken,hong-hanh.nguyen-le\}@ucdconnect.ie},
\email{an.lekhac@ucd.ie}}

\maketitle              
\begin{abstract}
Digital forensics faces unprecedented challenges with the emergence of digital twins and metaverse technologies. This paper presents the first comparative analysis between blockchain-based and traditional database systems for managing digital twin evidence in forensic investigations. We conducted controlled experiments comparing the Ethereum blockchain with IPFS storage against traditional SQL databases for digital twin evidence management. Our findings reveal that while blockchain provides superior data integrity and immutability, crucial for forensic applications, traditional databases offer better performance consistency. The blockchain implementation showed faster average storage times but higher variability in retrieval operations. Both systems maintained forensic integrity through hash verification, though blockchain's immutable nature provides additional security guarantees essential for legal proceedings. This research contributes to the development of robust digital forensic methodologies for emerging technologies in the metaverse era.

\end{abstract}
\section{Introduction}
The rapid evolution of digital twins and blockchain technologies has fundamentally transformed the landscape of criminal investigations and digital forensics. Digital twins \cite{mihai2022digital}, sophisticated virtual replicas of physical entities, processes, or systems, enable comprehensive modeling and simulation of real-world scenarios through advanced sensor data integration and computational analysis. Blockchain technology \cite{zheng2018blockchain} is a distributed ledger that maintains cryptographically secured and immutable transaction records across a decentralized network. The emergence of these technologies \cite{suhail2022blockchain} presents a powerful tool for managing digital evidence and maintaining evidence integrity.  

In digital forensic contexts, digital twins can serve as comprehensive evidence repositories that capture not only static data but also dynamic interactions and environmental conditions \cite{becker2024role}. This capability has potential in legal proceedings, which can be used in the courtroom as a copy of the physical evidence and the possibility of making detailed examinations of complex scenarios. The integration of blockchain technology with digital twins promises to enhance evidence handling by providing a decentralized, immutable, and transparent ledger for recording interactions and metadata associated with digital twin evidence. Particularly, blockchain technology enables digital evidence to be encrypted and stored online with metadata in a private manner and avoid unauthorized access \cite{suhail2022blockchain,yaqoob2020blockchain}. However, the practical performance implications of using blockchain-based systems for managing potentially large and complex digital twin data, compared to traditional database systems \cite{silva2016sql}, remain underexplored in digital forensic contexts.

This paper addresses this gap by conducting a comparative analysis of \\ blockchain-based versus traditional SQL database systems for the storage and retrieval of digital twin evidence. To achieve this, we develop two parallel environments: a private Ethereum blockchain \cite{buterin2013ethereum} with InterPlanetary File System (IPFS) storage \cite{ipfsIPFSDocumentation} versus a traditional MySQL database system. We evaluate the efficiency, which is defined as the time taken for storage (POST) and retrieval (GET) operations, for digital twins of varying sizes (i.e., ranging from 1MB to 200MB) across both systems. Furthermore, we assessed the forensic utility of each approach, primarily by verifying the integrity of the evidence via hash value comparisons. A web interface using the A-Frame framework is also developed to enable interactive 3D visualization of digital twin evidence, supporting forensic analysis workflows. In experiments, we employ Cohen's d \cite{diener2010cohen} for effect size determination and linear regression for scalability assessment.

\textbf{Contribution.} Our contributions are summarized as follows:
\begin{itemize}
    \item \textbf{A quantitative performance comparison}: We first provide a quantitative comparison of blockchain-based and traditional SQL database systems, specifically for digital twin evidence management in forensic applications. 
    \item \textbf{An assessment of forensic integrity mechanisms}: We evaluate the forensic soundness of both systems by implementing and verifying hash-based integrity checks. This includes leveraging the immutable nature of blockchain for recording IPFS hashes and comparing them with hash storage and verification in a SQL database.
    \item \textbf{Implementation of experimental frameworks}: We provide a detailed design and implementation of two controlled experimental environments, including the integration of Ethereum, Solidity smart contracts, IPFS, a MySQL database, and a unified Python-based web application featuring an A-Frame interface for interaction and 3D visualization.
    \item \textbf{An analysis of scalability and performance-integrity trade-offs}: We offer insights into the scalability implications and performance-integrity trade-offs inherent in each system. This provides guidelines for practitioners and researchers when selecting or designing systems for digital twin evidence management.
\end{itemize}


\section{Related Works}
\subsection{Blockchain in Digital Forensics}
Recent research has extensively explored blockchain's potential for enhancing digital forensics capabilities. Li et al. \cite{li2019blockchain} developed a blockchain-based framework for Internet-of-Things (IoT) forensics, namely IoT-fog-cloud (IoTFC), which addresses challenges of evidence authenticity and traceability in dynamic environments. With this framework, the transparent audit trails are provided to enhance trust between investigative entities. Gopalan et al. \cite{gopalan2019digital} demonstrated how the decentralized and tamper-proof nature of blockchain can be adopted to assure maintenance of evidence integrity and purity. Their proposed blockchain-based system hashes the data and stores it in blocks, ensuring that digital evidence is kept free from unauthorized
changes or deletions. Another study has provided a detailed analysis of how blockchain technology can be applied to improve control over the chain of custody in forensic science, particularly physical evidence \cite{batista2023exploring}. Alqahtany and Syed et al. \cite{alqahtany2024forensictransmonitor} introduced ForensicTransMonitor, a comprehensive blockchain approach for forensic transaction and evidence preservation. Their system leverages blockchain's immutability to ensure evidence integrity while providing automated verification mechanisms through smart contracts. This means that all activities within the forensic process are recorded as verifiable transactions on the blockchain.

The application of blockchain for provenance management has shown promising results in scientific data management. Ramachandran and Kantarcioglu et \cite{ramachandran2017using} developed DataProv, which uses blockchain's distributed nature to prevent unauthorized modifications while implementing smart contracts for automated verification. Their system demonstrated effectiveness in tracking clinical drug trials and agricultural supply chains, highlighting blockchain's versatility in maintaining data integrity across various domains.

\subsection{Applications of Digital Twins}
Digital twins have emerged as powerful tools for industrial IoTs. Attaran et al. \cite{attaran2024digital} explored the relationship between digital twins and industrial IoT. The article identifies some applications of digital twins in manufacturing by developing exact replicas of physical systems, which can be used to monitor them in real-time and also for predictive maintenance. Artificial intelligence (AI) algorithms have been leveraged to improve the prediction or automation capabilities of the digital twins. A modular approach is introduced by Zhang et al. \cite{zhang2017digital} proposed an intelligent multi-objective optimization algorithm which allows the digital twin to not only simulate the production line but also to optimize its design and dynamic execution. Sun et al. \cite{sun2020digital} utilized models like CNNs, RNNs, and PCNNs to process and structure complex, multi-source data, including images, noise, and 3D point clouds from the assembly process. This approach helps enhance the assembly-commissioning process and autonomous decision-making. 

The integration of digital twins with cyber-physical-social systems has shown particular promise for security applications. Han et al \cite{han2022paradefender} introduced ParaDefender, a scenario-driven parallel architecture that uses digital twins to enhance metaverse security. Their system demonstrates how digital twins can be used for a real-world use case. 

\subsection{Metaverse Digital Forensics}
The emergence of metaverse technologies has created new forensic challenges requiring specialized tools and methodologies. Kim et al \cite{kim2023digital} outlined approaches for extracting and analyzing data from Head-Mounted Displays (HMDs). This work highlights the unique aspects of metaverse forensics where evidence exists both on physical devices and within virtual environments. Huynh-The et al. \cite{huynh2023blockchain} examined blockchain's role in metaverse security, emphasizing how distributed ledgers can support data integrity and non-repudiation in virtual environments. Their research identifies the convergence of Virtual Reality (VR) and blockchain technologies as creating new dimensions for forensic investigations, where digital interactions must be verifiable and auditable.

\textbf{Research gaps identified from previous works.} Research gaps identified in the literature include (i) the lack of comprehensive performance comparisons between blockchain and traditional database systems for forensic applications, (ii) limited evaluation of hybrid storage approaches, and (iii) insufficient investigation of blockchain's impact on forensic workflow efficiency.

\section{Problem Statement and Forensic Approach}
\subsection{Motivations}
Current evidence management systems in digital forensics rely heavily on centralized databases and traditional storage architectures that present significant vulnerabilities in terms of data integrity, chain of custody maintenance, and cross-jurisdictional evidence sharing. The centralized nature of these systems requires complex trust relationships between investigating agencies, particularly in international collaborations where evidence authenticity and tamper-proofing become critical legal considerations \cite{hofmeister2017icrc}. The integration of blockchain technology with digital twins in legal spheres represents opportunities:
\begin{itemize}
    \item A detailed and accurate representation of crime scenes or incidents can be done by digital twins, which will enable investigators to collect comprehensive forensic data. VR equipment can be used for immersive analysis, allowing forensic experts to walk through virtual environments with all the tiny details.
    \item The hash of digital twins is stored on a private Ethereum blockchain while the actual digital twin is stored on the IPFS network. This improves the security and integrity of the data because Blockchain will ensure that evidence has not been tampered with, while IPFS guarantees efficient as well as secure storage and retrieval of these files.
    \item Blockchain technology guarantees transparency in recording all transactions and changes that digital twins undergo.
\end{itemize}

\subsection{Forensic Methodology}
This work investigates the efficacy of blockchain-based and traditional SQL database systems in managing digital twin models through a quantitative comparative analysis, focusing on the efficiency of both systems. To achieve this analysis, two parallel controlled environments are implemented to systematically assess the performance, integrity, and forensic utility of each approach. 

In our methodology, digital twins are treated as forensic artifacts that may represent:
\begin{itemize}
    \item Crime scene reconstructions with spatial and temporal data
    \item 3D models of physical evidence (weapons, vehicles, or other objects)
    \item Environmental conditions and their evolution over time
    \item Virtual reconstructions of incidents based on multiple data sources
\end{itemize}

\begin{wrapfigure}[28]{r}{0.6\textwidth}
  \centering
  \includegraphics[width=0.58\textwidth]{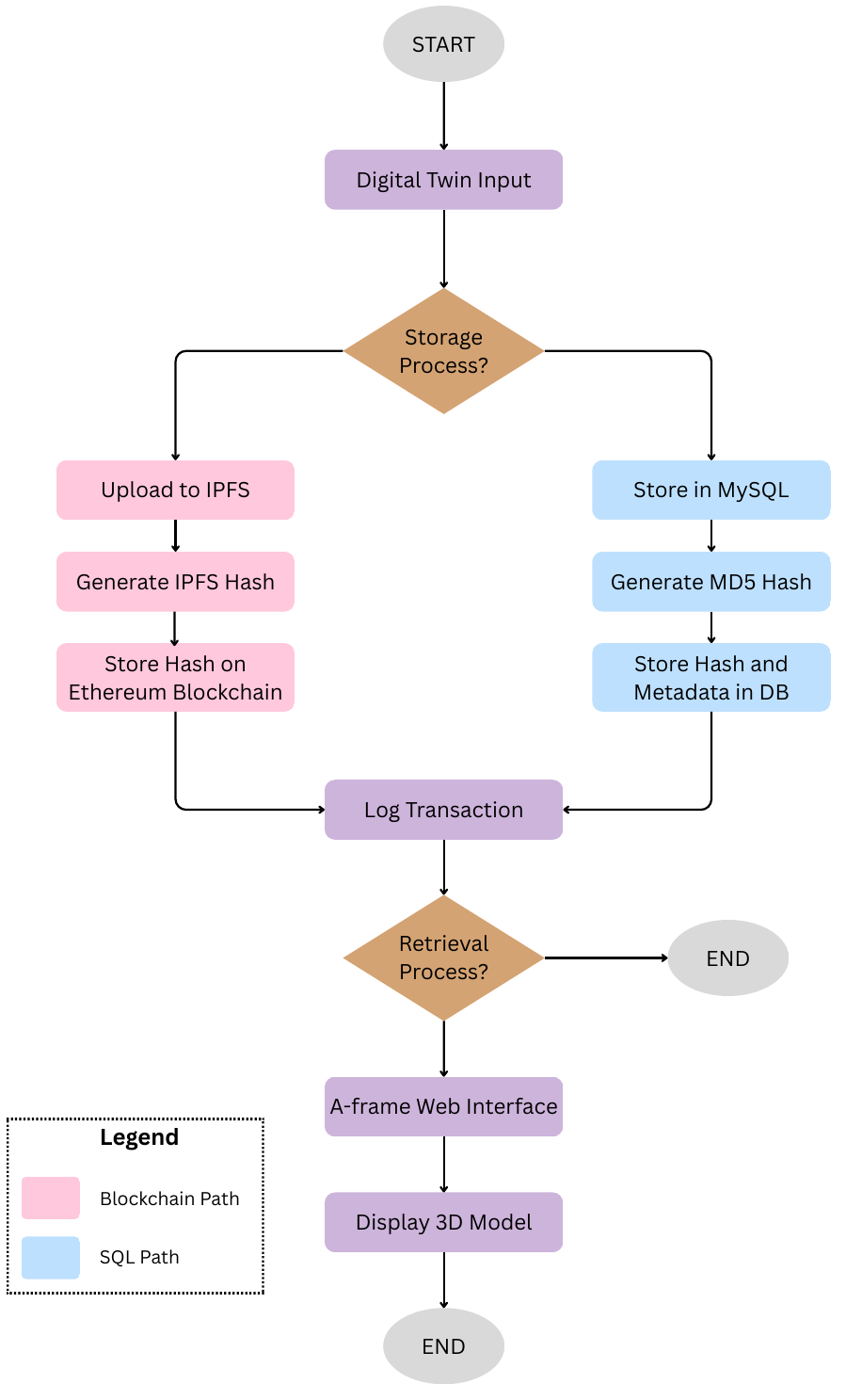}
  \caption{Flow chart of adopted forensic approach.}
  \label{fig:flow-chart}
\end{wrapfigure}

Each digital twin undergoes cryptographic hashing (MD5) upon submission to ensure integrity verification throughout the evidence lifecycle. The hash values are stored alongside the evidence in both systems, enabling forensic examiners to verify that digital twins have not been altered or corrupted during storage or transmission.

\subsubsection{Controlled Environments Implementation.} We establish two distinct forensic evidence management systems:

\textbf{Blockchain-based system}: A private Ethereum blockchain network \cite{buterin2013ethereum} integrated with the InterPlanetary File System (IPFS) \cite{ipfsIPFSDocumentation} for decentralized storage. Smart contracts, developed in Solidity \cite{wohrer2018smart}, manage the interaction with the blockchain. The digital twin files are stored on the IPFS via the Pinata.cloud pinning service \cite{pinataPinningService}. The core idea is to store the large digital twin files off-chain on IPFS and record only their immutable IPFS hash and associated metadata on the blockchain. 

\textbf{Traditional SQL database system}: A standard MySQL database is employed, which is hosted on the Aiven.com cloud service \cite{aivenAivenYour}. In this environment, digital twin files are stored directly as binary large objects (BLOBs) along with their associated metadata and MD5 hash values within the SQL database. 

Both environments are accessed through a unified Python-based backend that standardizes the storage and retrieval operations, ensuring fair comparison between the two approaches.

\subsubsection{Storage and Retrieval Processes.} 
Figure \ref{fig:flow-chart} illustrates two primary processes of our forensic methodology. 

\textbf{Storage Process (POST)}: For the blockchain system, when digital evidence in the form of a digital twin is submitted to the system,  it is uploaded to IPFS. The IPFS then returns a unique content identifier, which is then recorded on the Ethereum blockchain along with the file's MD5 hash through a smart contract transaction. Regarding the SQL system, the digital twin file is directly stored in the database along with its computed MD5 hash and associated metadata.

\textbf{Retrieval Process (GET)}: This process is conducted when forensic investigators need to access evidence. For the blockchain system, the smart contract is queried using the block number to retrieve the IPFS hash, which is then used to fetch the digital twin from the IPFS network. In contrast, a direct database query is performed in the SQL system to retrieve the digital twin based on its unique identifier.

\begin{figure}[ht]
    \centering
    \includegraphics[width=0.9\textwidth]{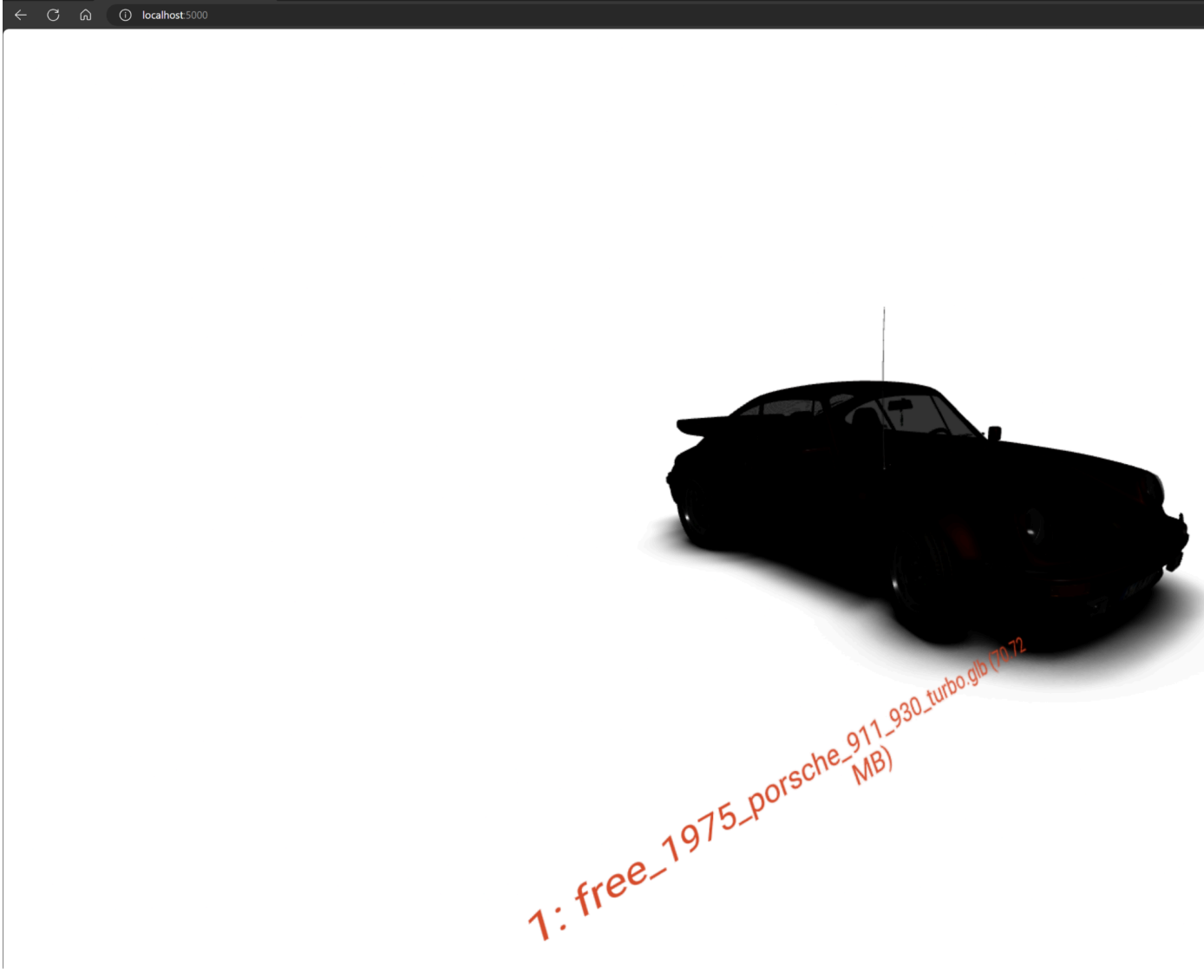}
    \caption{An example of the retrieved digital twin in the web application.}
    \label{fig:web}
\end{figure}

\subsubsection{Web Interface and Visualization Framework.} To facilitate forensic analysis and evidence examination, we develop a web-based interface using the A-Frame framework \cite{aframeAFrameMake}, which provides immersive 3D visualization capabilities essential for digital twin evidence analysis. There are two main purposes of this interface: (i) Investigators can upload digital twin models through a standardized web form, initiating the storage process in either the blockchain or SQL system; (ii) The interface provides an analysis environment for investigators to interact with 3D digital twin models in real-time, examine evidence from multiple angles, and compare evidence across different time periods or versions. Figure \ref{fig:web} illustrates the digital twin file shown on the web application. 

The web application maintains system-agnostic functionality, seamlessly retrieving and displaying digital twins regardless of whether they are stored via blockchain/IPFS or traditional SQL database. Moreover, it also ensures consistent forensic analysis capabilities across both storage paradigms.

\section{Experiments}
\subsection{Tools} Tools used to build the blockchain-based system, the traditional SQL database system, and the A-frame webpage are summarized in Table \ref{tab:tools}.

\subsection{Datasets} To conduct our analysis, we collect a dataset of digital twin models sourced from publicly available 3D model repositories. Because of the inherent sensitivity of actual case data, utilizing public sources provided an ethical means to acquire diverse models for testing. Our dataset consists of 10 distinct 3D models, selected to reflect potential digital twin evidence scenarios, such as vehicle reconstructions. These models are specifically selected to cover a range from 1MB to 200MB. 

\begin{table}[ht]
\centering
\fontsize{7pt}{7pt}\selectfont
\caption{Summarization of tools used in our experiments.}
\label{tab:tools}
\begin{tblr}{
  width = \linewidth,
  colspec = {Q[180]Q[192]Q[90]Q[433]},
  row{1} = {c},
  cell{2}{1} = {c},
  cell{2}{2} = {c},
  cell{2}{3} = {c},
  cell{3}{1} = {c},
  cell{3}{2} = {c},
  cell{3}{3} = {c},
  cell{4}{1} = {c},
  cell{4}{2} = {c},
  cell{4}{3} = {c},
  cell{5}{1} = {c},
  cell{5}{2} = {c},
  cell{5}{3} = {c},
  cell{6}{1} = {c},
  cell{6}{2} = {c},
  cell{6}{3} = {c},
  cell{7}{1} = {c},
  cell{7}{2} = {c},
  cell{7}{3} = {c},
  cell{8}{1} = {c},
  cell{8}{2} = {c},
  cell{8}{3} = {c},
  cell{9}{1} = {c},
  cell{9}{2} = {c},
  cell{9}{3} = {c},
  cell{10}{1} = {c},
  cell{10}{2} = {c},
  cell{10}{3} = {c},
  cell{11}{1} = {c},
  cell{11}{2} = {c},
  cell{11}{3} = {c},
  cell{12}{1} = {c},
  cell{12}{2} = {c},
  cell{12}{3} = {c},
  hlines,
  vline{2-4} = {-}{},
  hline{1,13} = {-}{0.08em},
  hline{2} = {-}{solid,black},
  hline{2} = {2}{-}{solid,black},
}
\textbf{Tool}                                 & \textbf{Category}                & \textbf{Version} & \textbf{Role/Purpose in the Experiment}                                        \\
Ganache \cite{trufflesuiteGanacheTruffle}     & Blockchain environment           & v2.7.1           & Served as the private blockchain system for for controlled testing environment \\
Pinata \cite{pinataPinningService}            & Blockchain environment           & API v1           & Decentralized IPFS file storage for digital twin models\textsuperscript{}      \\
Solidity \cite{soliditylangIntroductionSmart} & Blockchain environment           & v0.8.19          & Smart contract development for evidence management logic                       \\
MySQL~                                        & Database environment             & v8.0             & Relational database for traditional evidence storage                           \\
Aiven \cite{aivenAivenYour}                   & Database environment             & Custom           & Cloud database hosting service                                                 \\
PyMySQL                                       & Database environment             & v1.0.3           & Python MySQL database connector                                                \\
A-Frame                                       & User interface and visualization & v1.5.0           & Web-based VR framework for 3D model visualization                              \\
Flask                                         & User interface and visualization & v2.3.2           & Python web framework for application backend\textsuperscript{}                 \\
HTML5 / CSS3 / JavaScript                         & User interface and visualization & ES6+             & Frontend technologies for web interface                                        \\
Python                                        & Development and analysis tools   & v3.9+            & Primary programming language for system implementation                         \\
SPSS                                          & Development and analysis tools   & v29.0            & Statistical analysis and hypothesis testing                                    
\end{tblr}
\end{table}

\subsection{Configuration for Storage and Retrieval Operations}
\textbf{Ethereum Blockchain Configuration}: The blockchain environment utilized a private Ethereum network (Ganache) with the following specifications:
\begin{itemize}
    \item Network ID: Private testnet for controlled experimentation.
    \item Smart contract deployment using Solidity for evidence management.
    \item IPFS integration through Pinata service (1GB free tier).
    \item Web3.py library for blockchain interaction.
    \item JSON configuration file containing blockchain URLs, contract addresses, and API credentials.
\end{itemize}

\textbf{SQL Database Configuration}: The SQL environment employed MySQL with the following setup:
\begin{itemize}
    \item Aiven cloud hosting service (5GB free tier). The configuration includes host address, port number, authentication credentials, and database name.
    \item Direct file storage with metadata tables.
    \item PyMySQL driver for database connectivity.
    \item Additional parameters such as character set (utf8mb4) and timeout values for connection, read, and write operations are set to 10 seconds each to ensure reliable database operations.
\end{itemize}

\begin{table}[h!]
\centering
\fontsize{7pt}{7pt}\selectfont
\caption{Experimental variables and their characteristics.}
\label{tab:variables}
\begin{tblr}{
  width = \linewidth,
  colspec = {Q[192]Q[148]Q[348]Q[246]},
  cells = {c},
  hlines,
  vline{2-4} = {-}{},
  hline{1,7} = {-}{0.08em},
  hline{2} = {-}{solid,black},
  hline{2} = {2}{-}{solid,black},
}
\textbf{Variable}  & \textbf{Type}                & \textbf{Description}                & \textbf{Range/Values} \\
File size          & {Independent\\(Continuous)}  & Size of digital twin evidence files & 1-200 MB              \\
Storage method     & {Independent\\(Categorical)} & Type of storage system              & Blockchain/IPFS, SQL  \\
Internet speed     & {Independent\\(Continuous)}                   & Network bandwidth                   & 100/100 Mbps          \\
Execution time     & Dependent                    & Time for POST/GET operations        & Measured in seconds   \\
Forensic integrity & Dependent                    & Hash verification success           & Boolean (Pass/Fail)   
\end{tblr}
\end{table}

\subsection{Key Variables} 
Table \ref{tab:variables} summarizes the key variables in our experimental design. File size serves as an independent continuous variable ranging from 1 to 200 MB, representing various sizes of digital twin evidence files. Storage method is an independent categorical variable with two levels: blockchain with IPFS and traditional SQL database. Internet speed is a continuous independent variable, which is controlled at 100/100 Mbps to maintain consistent network conditions. Execution time, measured in seconds, serves as the primary dependent variable for both POST and GET operations. Forensic integrity is assessed as a binary dependent variable indicating successful hash verification.






\subsection{Metrics for Evaluation}
To evaluate the performance and forensic utility of both the blockchain-based and traditional SQL database systems for digital twin evidence management, we employ a set of quantitative and qualitative metrics. 

\textbf{Efficiency analysis.} The primary metric for efficiency is the execution time (in seconds) required for both storage (POST) and retrieval (GET) operations. For each transaction, we calculate the mean execution time:
\begin{equation}
    \bar{T} = \frac{1}{n} \sum^n_{i=1}T_i,
\end{equation}
where $n$ is the number of files and $T_i$ is the execution time for the $i$-th transaction: $T_i = T_{i_{\text{end}}} - T_{i_{\text{start}}}$. 

To quantify the magnitude of the difference in performance between the two systems, we calculated Cohen's D \cite{diener2010cohen}, which is computed as follows:
\begin{equation}
    D = \frac{\bar{T}_{\text{blockchain}} - \bar{T}_{\text{SQL}}}{\sigma_{\text{pooled}}},
\end{equation}
where $\bar{T}_{\text{blockchain}}$ and $\bar{T}_{\text{SQL}}$ are the means of execution time of blockchain-based and traditional SQL database systems, respectively; and $\sigma_{\text{pooled}}$ is the pooled standard deviation, which is computed as
$\sigma_{\text{pooled}} = \frac{\sigma^2_{\text{blockchain}} + \sigma^2_{\text{SQL}}}{2}$. This effect size measure helps in understanding the practical significance of the observed differences beyond statistical significance.

\textbf{Scalability analysis.} To assess how each system performs as the size of the digital twin files increases, we conducted a \textit{linear regression analysis} \cite{devore2003linear}. This helps in understanding the relationship between file size (independent variable) and execution time (dependent variable). We fit separate regression models for each storage method and operation type. The general model is:
\begin{equation}
    \text{ExecutionTime} = \beta_0 + \beta_1 \times \text{FileSize} + \epsilon,
\end{equation}
where $\beta_0$ is the intercept, $\beta_1$ is the coefficient representing the change in time per unit increase in file size, and $\epsilon$ is the error term. 

\textbf{Forensic integrity assessment.} Forensic integrity is evaluated as a binary outcome (Pass/Fail) based on hash value verification. Upon retrieval, the MD5 hash of the digital twin file is recalculated and compared against the hash value stored during the initial storage operation. 

\section{Description of Results}
In this section, we present results and findings from our comparative experiments. We focus on the efficiency and forensic integrity of managing digital twin evidence using both the blockchain-based and traditional SQL database systems. 

\subsection{Descriptive Statistics} 

We conduct controlled experiments to analyze the execution times for storing and retrieving 10 digital twin models ($n=10$) of varying sizes (i.e., ranging from 1MB to 200MB). The experimental results from both blockchain and traditional SQL database systems reveal significant performance differences across storage and retrieval operations. Table \ref{tab:descriptive} presents the descriptive statistics for both storage (POST) and retrieval (GET) operations across blockchain-based and traditional SQL database systems.
 
\begin{table}[ht]
\centering
\fontsize{8pt}{8pt}\selectfont
\caption{Descriptive statistics of experimental results.}
\label{tab:descriptive}
\begin{tblr}{
  width = \linewidth,
  colspec = {Q[175]Q[173]Q[160]Q[129]Q[146]Q[146]},
  cells = {c},
  hlines,
  vline{2-6} = {-}{},
  hline{1,6} = {-}{0.08em},
  hline{2} = {-}{solid,black},
  hline{2} = {2}{-}{solid,black},
}
Operation Type  & Storage Method & Mean Time (s) & Std Dev (s) & Min Time (s)~ & Max Time (s) \\
Storage (POST)  & Blockchain     & 16.43         & 9.43        & 4.08          & 30.97        \\
Storage (POST)  & SQL Database   & 25.36         & 16.85       & 2.56          & 51.77        \\
Retrieval (GET) & Blockchain     & 15.95         & 10.65       & 3.36          & 23.59        \\
Retrieval (GET) & SQL Database   & 11.01         & 7.21        & 1.13          & 22.59        
\end{tblr}
\end{table}

From Table \ref{tab:descriptive}, we can observe that blockchain systems achieved faster average performance for storage operations ($16.43$s vs $25.36$s) while SQL database outperforms the blockchain system ($11.01$s vs $15.95$s) in terms of retrieval operations. 

\begin{figure}[h!]
    \centering
    \begin{subfigure}[t]{0.48\textwidth}
        \centering 
        \includegraphics[width=\textwidth]{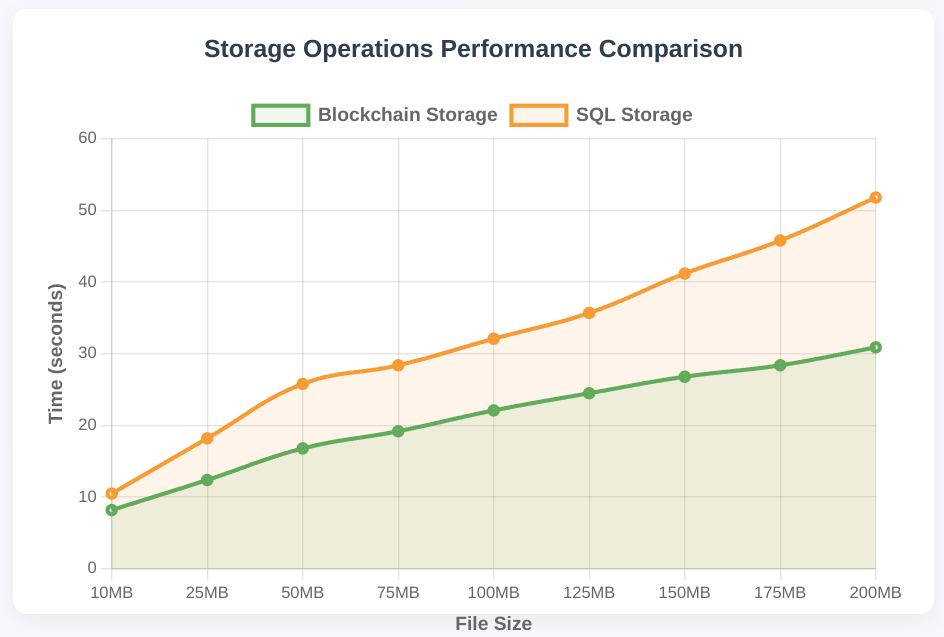}
        \caption{Storage Operations.}
        \label{fig:storage}
    \end{subfigure}
    \hfill
    \begin{subfigure}[t]{0.48\textwidth}
        \centering 
        \includegraphics[width=\textwidth]{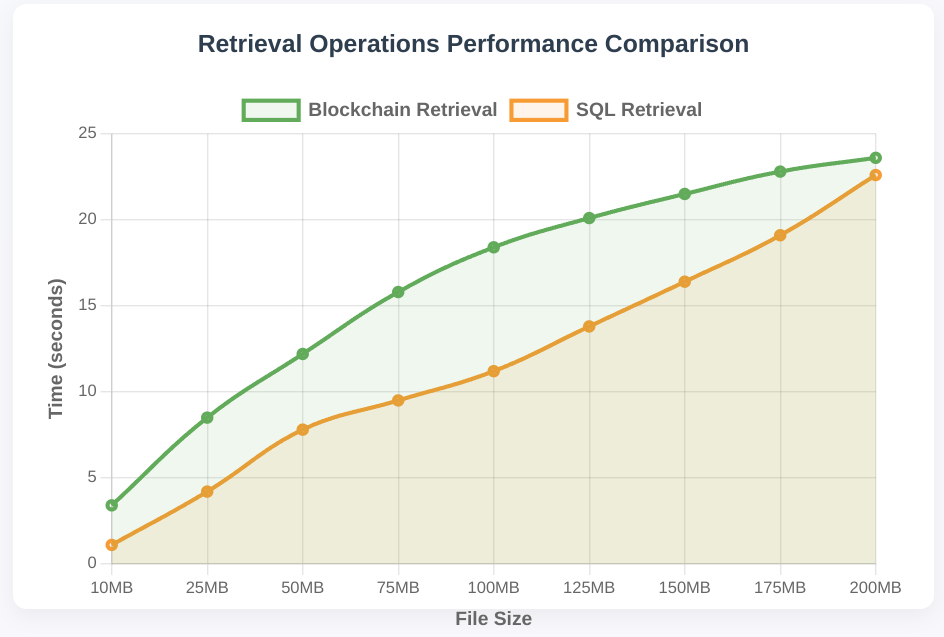}
        \caption{Retrieval Operations.}
        \label{fig:retrieval}
    \end{subfigure}

    \caption{Comparison of storage and retrieval operations of blockchain-based and SQL database systems.}
    \label{fig:performance}
\end{figure}

\subsection{Performance Analysis by File Size}
We analyzed the execution times for storing and retrieving 10 digital twin models on two systems. Figure \ref{fig:storage} illustrates the relationship between file size and storage execution time for both systems. The blockchain-based system demonstrates superior storage performance with $35\%$ faster average times compared to SQL databases. This result suggests that the blockchain/IPFS hybrid architecture benefits from batch processing efficiencies and distributed storage mechanisms when handling larger files.

In contrast, regarding retrieval operations, Figure \ref{fig:retrieval} reveals that SQL databases maintain more consistent performance across varying file sizes, with $31\%$ faster average times. However, when the file size increases to 200MB, the difference in retrieval execution time between the two systems becomes lower. This indicates that the blockchain-based system might be effective for larger files compared to the SQL database system. 

\subsection{Scalability Analysis.} 

To understand how each storage system handles increasing data loads, we analyzed the relationship between digital twin file size and execution time using linear regression. 

\begin{figure}[h!]
    \centering
    \begin{subfigure}[t]{0.48\textwidth}
        \centering 
        \includegraphics[width=\textwidth]{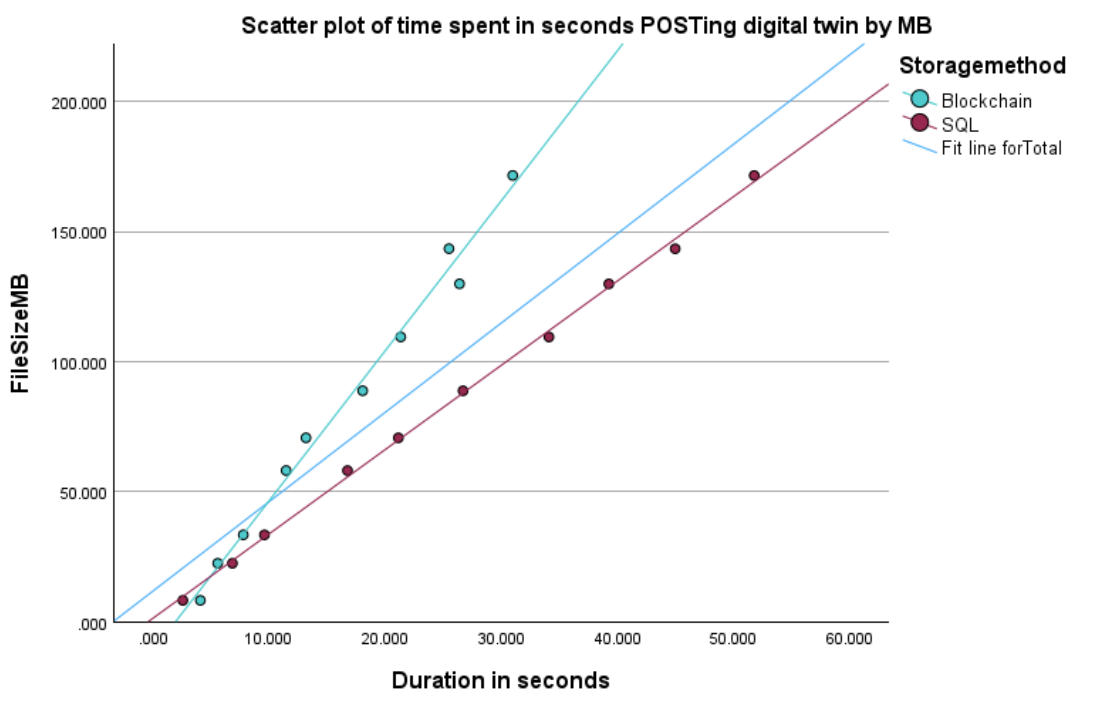}
        \caption{Storage Operations.}
        \label{fig:storage-scatter}
    \end{subfigure}
    \hfill
    \begin{subfigure}[t]{0.48\textwidth}
        \centering 
        \includegraphics[width=\textwidth]{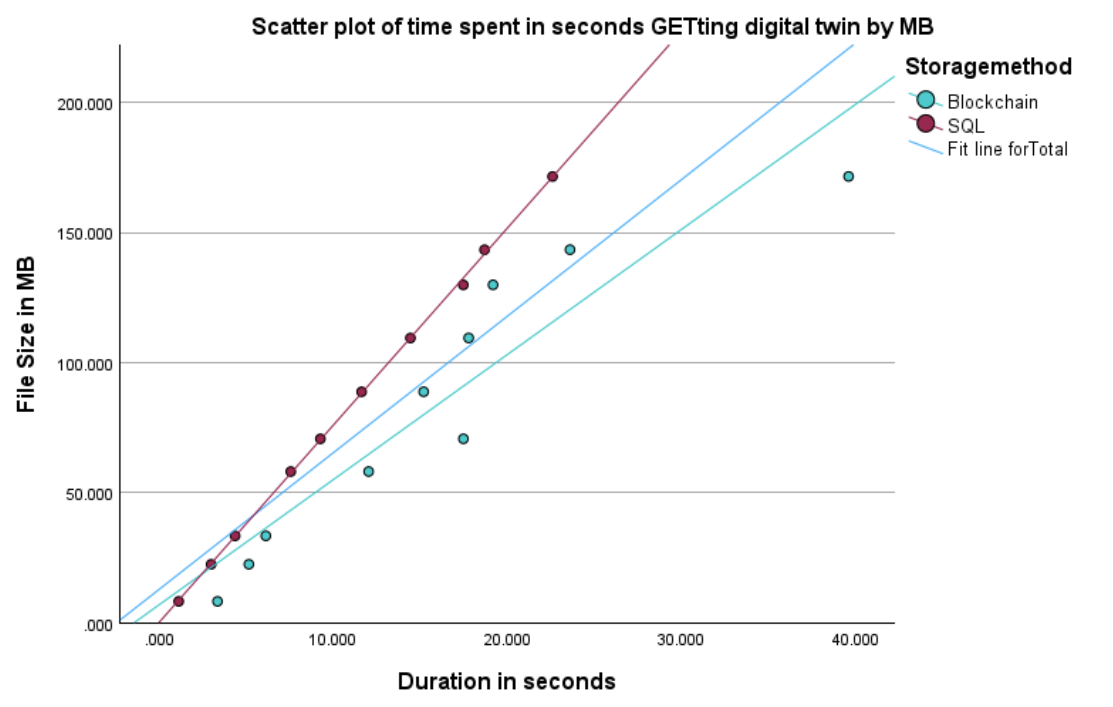}
        \caption{Retrieval Operations.}
        \label{fig:retrieval-scatter}
    \end{subfigure}

    \caption{Scatter plots of times spent for storage and retrieval operations of lockchain-based and SQL database systems.}
    \label{fig:scatter}
\end{figure}

The regression analysis (illustrated in Figure \ref{fig:scatter}) reveals important scalability characteristics:
\begin{itemize}
    \item \textbf{Storage Scalability} (Figure \ref{fig:storage-scatter}): While the average storage time of the SQL database system is higher, the trend line shows a more linear increase in time with file size. Although exhibiting more variability, the trend line for blockchain storage is less steep than SQL's. This suggests that as file sizes grow, the increase in storage time is less pronounced for the blockchain system, implying potentially better scalability for storage operations.
    \item \textbf{Retrieval Scalability} (Figure \ref{fig:retrieval-scatter}): The trend line for SQL retrieval was less steep and more linear. This indicates that SQL retrieval times increase at a slower and more predictable rate as the file size grows, suggesting better scalability for retrieval operations. In contrast, the time required for retrieval operations of the blockchain system grows more rapidly with larger file sizes, indicating potential scalability challenges for handling bigger datasets during retrieval. 
\end{itemize}

\subsection{Forensic Integrity Analysis}
We assess the data integrity of two systems by verifying hash values. During the evaluation, we compare the hash values stored in both systems against the original MD5 hashes generated when the data was first created. The result demonstrates that both systems successfully maintained forensic soundness, as all hash values correctly match the originals. This confirms that both systems can provide a verifiable record that the digital twin evidence is not altered. However, the inherent immutability and decentralized nature of the blockchain provide an additional layer of security and trust, which is particularly valuable for legal proceedings where proving non-tampering is essential.

\section{Discussion of Results}
Our comparative analysis between blockchain-based (Ethereum/IPFS) and traditional SQL database systems for managing digital twin evidence has revealed significant insights into their respective strengths and weaknesses within a digital forensics context. The results highlight a crucial trade-off between performance efficiency and the robustness of data integrity guarantees. 

First, our experiments show that the blockchain-based system is faster for storing digital twin evidence, suggesting efficiency in handling large files. However, the SQL database system outperforms for the retrieval operation. 

Second, the results indicate a scalability trade-off: SQL appears more scalable for consistent and efficient retrieval of digital twins, while the blockchain/IPFS approach shows promise for more scalable storage in this specific experiment, despite its inherent performance variability.

Crucially, while both systems maintain forensic integrity through hash verification, blockchain's inherent immutability and decentralization provide a significantly higher level of security. This is vital for legal proceedings where proving evidence has not been tampered with is essential.

Therefore, the choice depends on priorities. For rapid and consistent access, SQL remains strong. For cases demanding the highest level of verifiable integrity and trust, blockchain offers compelling advantages, despite its current performance overheads.

\section{Conclusion}
This work provides the first comparative analysis of blockchain-based and traditional database systems for digital twin evidence management in forensic contexts. Through controlled experiments, we evaluated performance based on storage (POST) and retrieval (GET) times for files ranging from 1MB to 200MB, alongside a crucial assessment of forensic integrity. Our findings underscore performance-integrity trade-offs: (i) The blockchain-based system is faster for the storage operation, while the SQL database system outperforms for the retrieval operation; and (ii) The inherent immutability and decentralized trust model of blockchain offer higher security guarantees. Future work should focus on optimizing blockchain performance and exploring hybrid models of blockchain and SQL database systems to harness the best of both paradigms.

%
%
%
\newpage
\section*{Acknowledgement}
\thanks{Corresponding author}
This research is partially funded by UCD ASEADOS Lab and UCD School of Computer Science.

\bibliographystyle{splncs04}
\bibliography{mybib}

\end{document}